\documentclass[fleqn,usenatbib]{mnras}
\usepackage{newtxtext,newtxmath}
\usepackage[T1]{fontenc}
\usepackage{ae,aecompl}
\usepackage{tabularx}
\usepackage{graphicx}
\usepackage{float}
\usepackage{mathrsfs}	
\usepackage{amsmath}
\usepackage{color}
\usepackage{hyperref}
\usepackage{multirow}
\usepackage{subfigure}
\usepackage{longtable}

\usepackage{graphicx}	% Including figure files
\usepackage{amsmath}	% Advanced maths commands
\usepackage{amssymb}	% Extra maths symbols
\usepackage{csvsimple}

\title{Constraining the Rastall parameters in static spacetimes with galaxy-scale strong gravitational lensing} 
\author[Li et al.]
{Rui Li$^{1,2,3,4}$\thanks{E-mail: lirui@ynao.ac.cn}, Jiancheng Wang$^{1,2,3,4}$\thanks{E-mail: jcwang@ynao.ac.cn}, Zhaoyi Xu$^{5}$, Xiaotong Guo$^{6}$
\vspace*{0.2cm}\\
$^{1}$ Yunnan Observatories, Chinese Academy of Sciences, 396 Yangfangwang, Guandu District, Kunming, 650216, P. R. China\\ 
$^{2}$ University of Chinese Academy of Sciences, Beijing, 100049, P. R. China \\
$^{3}$Center for Astronomical Mega-Science, Chinese Academy of Sciences, 20A Datun Road, Chaoyang District, Beijing, 100012,\\ 
\ \ \ P. R. China \\ 
$^{4}$ Key Laboratory for the Structure and Evolution of Celestial Objects, Chinese Academy of Sciences, 396 Yangfangwang, Guandu \\
\ \ \ District, Kunming, 650216, P. R. China\\
$^{5}$ Key Laboratory of Particle Astrophysics,
Institute of High Energy Physics, Chinese Academy of Sciences, Beijing 100049, China\\
$^{6}$ School of Astronomy and Space Science, Nanjing University, Nanjing, Jiangsu 210093, China}

\date{Accepted xxx. Received xxx; in original form xxx}

\pubyear{2018}

\begin{document}
\label{firstpage}
\pagerange{\pageref{firstpage}--\pageref{lastpage}}
\maketitle

% Abstract of the paper
\begin{abstract}
Recently, Rastall gravity is undergoing a significant surge in popularity. We obtain a power-law total mass-density profile for the inner region (within several effective radius) of early-type galaxies (ETGs) from the space-time structures which are described by the static spherically-symmetric solutions of Rastall gravity under the assumption of perfect fluid matter. We find that in the inner region of ETGs, the Rastall dimensionless parameter $\beta=\kappa\lambda$ determines the mass distribution. We then use 118 galaxy-galaxy strong gravitational lensing systems to constrain the Rastall dimensionless parameter $\beta$. We find that the mean value of $\beta$ for total 118 ETGs is $\beta=0.163\pm0.001$(68\% CL) with a minor intrinsic scatter of $\delta=0.020\pm 0.001$.  Our work observationally illustrates the physical meaning of  the Rastall dimensionless parameter in galaxy scale. From the Newtonian approximation of Rastall gravity, we also find that an absolute isothermal mass distribution for ETGs is not allowed in the framework of Rastall gravity.
\end{abstract}

% Don't make up new ones.

%%%%%%%%%%%%%%%%% BODY OF PAPER %%%%%%%%%%%%%%%%%%
\section{Introduction}
The energy-momentum conservation law (OCL) is one of the basic elements in General Relativity (GR), but as \cite{1972PhRvD...6.3357R} argued, it is fuzzy and unclear. Therefore, some generalized GR have been put forward, aiming to relax the conservation condition of energy-momentum tensor. Rastall gravity \citep{1972PhRvD...6.3357R, 1976CaJPh..54...66R} is one of these generalized GR. In Rastall gravity, the OCL is changed to $T^{\mu\nu}_{~~;\mu}=\lambda R^{,\nu}$, and if the space-time is flat, the OCL would back to Einstein formalism. We can understand Rastall gravity under Mach principle which means that the inertia of a mass distribution is dependent on the mass and energy content of the external spacetime \citep{2006gr.qc....10070M}. For Rastall gravity, there are many applications in cosmology and black hole space-time \citep{2012PhRvD..85h4008B, 2018EPJP..133..249D, 2012IJMPS..18...67F, 2017PhLB..771..365H, 2017CaJPh..95.1253H, 2018IJMPD..2750069L}. But in galaxies, we have not found its applications. Also, there are no observational constrains to the parameters in Rastall gravity. %Recently, \cite{2018PhLB..782...83V} pointed out that Rastall gravity is equivalent to Einstein GR, the Rastall parameter $\kappa\lambda$ only re-arrange the matter sector of Einstein GR, which may have some applications to study the mass distribution of galaxies.

If one want to use Rastall gravity to study galaxies (e.g., the mass distribution), solving the field equation is necessary. However, the matter in galaxies is varied, bringing difficulties to solve the field equation. A practical alternative is to treat all the matter in galaxies as perfect fluid. As we know, a galaxy mainly contains two fractions, the stellar fraction and the dark matter fraction. For the stellar fraction, there are little sticky, shear stresses and viscosity. Therefore, we can safely treat the stellar fraction as perfect fluid. For the dark matter fraction, we know little about it. Many models of dark matter have been proposed. Among them, the cold dark matter model \citep[CMD;][]{1991ApJ...378..496D, 1996ApJ...462..563N} is the most popular one. However, at small scale (e.g., galaxy scale), several observations are not in accordance with the predictions of CDM, such as the ``missing satellites" problem and the ``cusp-core" controversy \citep[e.g.,][]{2009ASPC..419..283K, 2018LRR....21....2A}. Therefore, some other dark matter models, such as the warm dark matter model \citep[WDM;][]{2005PhRvD..71f3534V}, the fuzzy dark matter model \citep[FDM;][]{1990PhRvL..64.1084P, 1994PhRvD..50.3650S}, have been proposed to explain the observations. Recently, a perfect fluid dark matter model \citep[PFDM;][]{2010PhLB..694...10R, 2011IJTP...50.2655R} was also proposed, in which the dark matter is described as perfect fluid without sticky, shear stresses, viscosity. \cite{2010PhLB..694...10R} explained that many dark matter models predict anisotropic dark matter fluid stress tensor when explain the flat rotation curves of spiral galaxies. However, no physical mechanisms and observational evidences support it. Therefore, treating dark matter as an isotropic perfect fluid seems to be reasonable, because the predictions from such model at stellar and cosmic scales have been identified by observations.

In this work, we use galaxy-scale strong gravitational lensing to constrain the parameters in Rastall gravity. We know that the light can be blended by massive objects. If a source galaxy is located behind a foreground galaxy, its light would be blended by the gravity of the foreground galaxy. Therefore,  around the foreground galaxy, we are able to find several images of the source galaxy. We call this phenomenon as galaxy-galaxy strong gravitational lensing. Generally, the locations of the images can help us to obtain the Einstein radius of the lensing system, and then we can obtain the precise Einstein mass (the total mass within the Einstein radius) from the Einstein radius. However, if we want to infer the total mass inside a sphere with any radius $r$ from the Einstein mass, the mass distribution of the foreground galaxy is also needed. For early-type-galaxies (ETGs), a power-law mass density profile is often assumed. For example, a method named the joint analysis of gravitational lensing and dynamical data is often used to study the mass properties of the foreground ETGs in lensing systems \citep{2006ApJ...649..599K, 2009ApJ...703L..51K, 2010ApJ...724..511A, 2012ApJ...757...82B, 2013ApJ...777...98S}. In this analytic technique, the power-law mass density profile is a basic assumption. Another example is about the cosmology. In order to obtain the accurate time-delay distances from strong lensing, \cite{2013ApJ...766...70S} used elliptically symmetric distributions with power-law profile to model the dimensionless surface mass density of the lens galaxies. The power-law mass density profile seems to be reasonable, as illustrated by the modeling of X-ray data \citep{2010MNRAS.403.2143H} and strong lensing and stellar kinematics \citep{2006ApJ...649..599K, 2010ApJ...724..511A}, but the origin of this profile remains unclear.

In this work, we derive a power-law mass density profile for ETGs from Rastall gravity by treating the matter in ETGs as perfect fluid. Then we use 118 gravitational lensing systems to constrain the parameters in Rastall gravity. Throughout this paper, $R$ is the radial coordinate in two-dimensions and $r$ is the radial coordinate in three-dimensions. We adopt a fiducial cosmological model with $\rm \Omega_m = 0.274$, $\rm \Omega_{\Lambda} = 0.726$, and $H_0 \rm = 70\,km\,s^{-1}\,Mpc^{-1}$.

\section{Theoretical basis} 
In this section, we will introduce the theoretical basis of our study, including the thoery of Rastall gravity, the spherical symmetry solution of Rastall gravity for perfect fluid matter and the knowledge of galaxy-galaxy strong gravitational lensing.

\subsection{Rastall gravity}
Rastall gravity was proposed in \cite{1972PhRvD...6.3357R}. In his work, Rastall argued that the ordinary OCL is not necessarily conserved in curved spactime. Because people have only checked the OCL in the flat spacetimes, we should not confine ourselves to it in the curved spacetime. Rastall also suggested that in general we should have $T^{\mu\nu}_{;\mu}\neq 0$, and such claim is in full agreement with recent observations \citep{2017PhRvL.118b1102J}. Rastall proposed to consider the OCL as
\begin{equation}
T^{\mu\nu}_{;\mu}=\lambda R^{,\nu}.
\end{equation}
This option leads the gravity field equation to be
\begin{equation}
G_{\mu\nu}+\kappa\lambda g_{\mu\nu}R=\kappa T_{\mu\nu},
\label{equ:rastall_equation}
\end{equation}
where $T_{\mu\nu}$ is the energy-momentum tensor of matter.  $\kappa$ and $\lambda$ are the Rastall gravitational coupling constant and the Rastall constant parameter, respectively. When $\lambda=0$, the gravity field Equations \ref{equ:rastall_equation} reduces to Einstein equation in GR. Rastall gravity does not introduce either a new energy-momentum tensor or a rearranged energy-momentum tensor, and it only attribute a new property to the energy-momentum tensor in the curved spacetime \citep{2018EPJC...78...25D}. Therefore, in Rastall gravity, there is an unknown interaction between geometry and matter field which needs more studies to be known and understood.

A true modified gravity should recover the Newtonian gravity at its appropriate limit. For Rastall gravity, the Newtonian approximation is \citep{1972PhRvD...6.3357R, 2017CaJPh..95.1257M, 2018EPJC...78...25D}
\begin{equation}
\frac{\kappa}{4\kappa\lambda -1}(3\kappa\lambda -\frac{1}{2})=\kappa_G,
\label{equ:Newton_limit}
\end{equation}
where $\kappa_G=4\pi G$ and $G$ is the Newton gravitational constant. Only when $\lambda=0$, the Einstein coupling constant  $\kappa_E=8\pi G$ is recovered. This equation will be used to infer the values of $\kappa$ and $\lambda$ in Section \ref{Sec:constrain}.

\subsection{Spherical symmetry solution of Rastall gravity for the perfect fluid}
In \cite{2017PhLB..771..365H}, the spherically symmetric space-time metric in the perfect fluid is given by
\begin{equation}
ds^{2}=-f(r)dt^{2}+f^{-1}(r)dr^{2}+r^{2}(d\theta^{2}+sin^{2}\theta d\phi^{2}).
\label{equ:matric}
\end{equation}
For Rastall gravity, $f(r)$ is
\begin{equation}
f(r)=1-\alpha r^{-\dfrac{1+3\omega-6\kappa\lambda(1+\omega)}{1-3\kappa\lambda(1+\omega)}},
\end{equation}
where $\alpha$ is an integration constant representing the surrounding field structures. $\omega$ is the equation of state defined by $\omega=p/\rho$, where $p$ and $\rho$ are the pressure and energy density of the perfect fluid, respectively. Substituting Equation \ref{equ:matric} into Equation \ref{equ:rastall_equation}, we can obtain the energy density of the perfect fluid matter
\begin{equation}
\begin{split}
\kappa\cdot(-\rho) &= f(r)[\dfrac{1}{r}\dfrac{f^{'}(r)}{f(r)}+\dfrac{1}{r^{2}}]-\dfrac{1}{r^{2}} \\
              &\ \ \ \ \ \ \ \ \ \ \ \ \ -\frac{1}{r^2}\kappa\lambda(r^2f''+4rf'-2+2f)\\
              &=-3\alpha\frac{(1-4\kappa\lambda)[\kappa\lambda(1+\omega)-\omega]}{[1-3\kappa\lambda(1+\omega)]^2} \\
              &\ \ \ \ \ \ \ \ \ \ \ \ \times r^{-\dfrac{3+3\omega-12\kappa\lambda(1+\omega)}{1-3\kappa\lambda(1+\omega)}}.
\end{split}
\label{Equ:density}
\end{equation}

In this work, we consider the case of galaxies. Because the velocity dispersion of the galaxies is much smaller than the speed of light, the energy density of the perfect fluid can be approximated as the mass density of the galaxies. Therefore, the total mass of the galaxies inside a sphere of radius $r$ is
\begin{equation}
M(r)=4\pi\int r^{2}\rho dr.
\end{equation}
Let's define $\beta=\kappa\lambda$, $\beta$ is a dimensionless constant, we call it as Rastall dimensionless parameter. We then obtain
\begin{equation}
M(r) =\frac{\alpha}{2\kappa} \frac{1-4\beta}{1-3\beta(1+\omega)}r^{\frac{3\beta(1+\omega)-3\omega}{1-3\beta(1+\omega)}},
\label{equ:theory_M}
\end{equation}
which describes how mass evolves as a function of radius $r$ on the basis of Rastall gravity. For galaxies and galaxy clusters, the evidences from theories and observations provide that $\omega$ is extremely close to 0 \citep[e.g.,][]{2010PhLB..694...10R, 2014ApJ...783L..11S, 2016PhLB..753..140P}. Therefore, we let $\omega=0$ in this work. Specially, if $\beta=0$, Equation \ref{equ:theory_M} reduces to GR formalism. Then the power-law index only depends on $\omega$. The mass of a galaxy would concentrate in the center (because $\omega$ is extremely close to 0),  meaning that GR can't describe the mass distribution of a galaxy.

%Here, we use the system of natural units, if we want to use the International System of Units, the $\alpha$ should be $\alpha c^2$, where $c$ is the speed of light. 

\subsection{Galaxy-galaxy strong gravitational lensing}
For a galaxy-galaxy strong gravitational lensing system, the light from the source galaxy could be deflected by the foreground galaxy, forming arcs or multiple images around the foreground galaxy. We can infer the Einstein radius $\theta_E$ through the locations of these arcs or images. The density inside the Einstein radius is the critical projected mass density described as
\begin{equation}
\Sigma_{crit}=\frac{c^2}{4\pi G}\frac{D_s}{D_lD_{ls}},
\end{equation}
where $D_l$ and $D_s$ are the angular-diameter distances of the lens and the source, respectively. $D_{ls}$ is the angular-diameter distance between the lens and the source. The mass inside the Einstein radius is named Einstein mass written as $M_{ein}=\pi R_E^2 \Sigma_{crit}$, where $R_E=D_l \theta_E$.

An empirical power-law total mass-density profile is usually assumed to describe the 3-dimensional mass distribution of ETGs \citep[e.g., ][]{2006ApJ...649..599K,2009ApJ...703L..51K,2010ApJ...724..511A,2012ApJ...757...82B}. The empirical total mass-density profile has a form of
\begin{equation}
\rho_{tot}=\rho_{0}(\frac{r}{r_0})^{-\gamma},
\label{equ:density}
\end{equation}
where $\gamma$ is the effective slope given by \cite{2004ApJ...611..739T}, $\rho_{0}$ is determined by the Einstein mass $M_{ein}$, and $r_0 $ can be chosen arbitrarily \citep{2006ApJ...649..599K}. For a lens galaxy, when the power-law total mass-density profile is assumed, the 3-dimensional total mass inside a sphere of radius $r$ can be obtained by projecting the 2-dimensional Einstein mass to the 3-dimensional space as
\begin{equation}
M(r)=f(\gamma)\cdot M_{ein}\cdot R_{ein}^{\gamma-3}\cdot r^{3-\gamma},
\label{equ:observe_M}
\end{equation}
where
\begin{equation}
f(\gamma)=\frac{2\Gamma(\frac{1}{2}\gamma)}{\Gamma(\frac{1}{2}\gamma-\frac{1}{2})\Gamma(\frac{1}{2})}.
\end{equation}

\section{Constraining Rastall parameters with gravitational lensing}\label{Sec:constrain}
Equation \ref {equ:theory_M} theoretically predicts the 3 dimensional total mass inside a sphere of radius $r$ in Rastall gravity, while Equation \ref{equ:observe_M} empirically describes the 3 dimensional total mass based on the assumption of a power-law mass-density profile. It is noted that Equation \ref {equ:theory_M} and Equation \ref{equ:observe_M} have a similar power-law form. Now, equaling the power-law indexes of  these two equations, we get
\begin{equation}
3-\gamma=\frac{3\beta(1+\omega)-3\omega}{1-3\beta(1+\omega)}.
\end{equation}
Assuming $\omega=0$, we obtain
\begin{equation}
\beta=\frac{3-\gamma}{12-3\gamma},
\label{Equ:gamma_beta}
\end{equation}
where $\gamma$ can be inferred from gravitational lensing. After known $\gamma$, we can easily obtain the values of $\beta$. 

%\begin{equation}
%\kappa=\frac{4\beta-1}{6\beta-1}\kappa_E,    \ \ \ \ \ \ \ \ \ \    \lambda=\frac{\kappa-2\kappa_G}{6\kappa^2-8\kappa %z%\kappa_G}
%\end{equation}

\begin{figure*}
\centering
\subfigure[]{\label{Fig.beta_distri} \includegraphics[height=0.3\textwidth, width=0.32\textwidth]{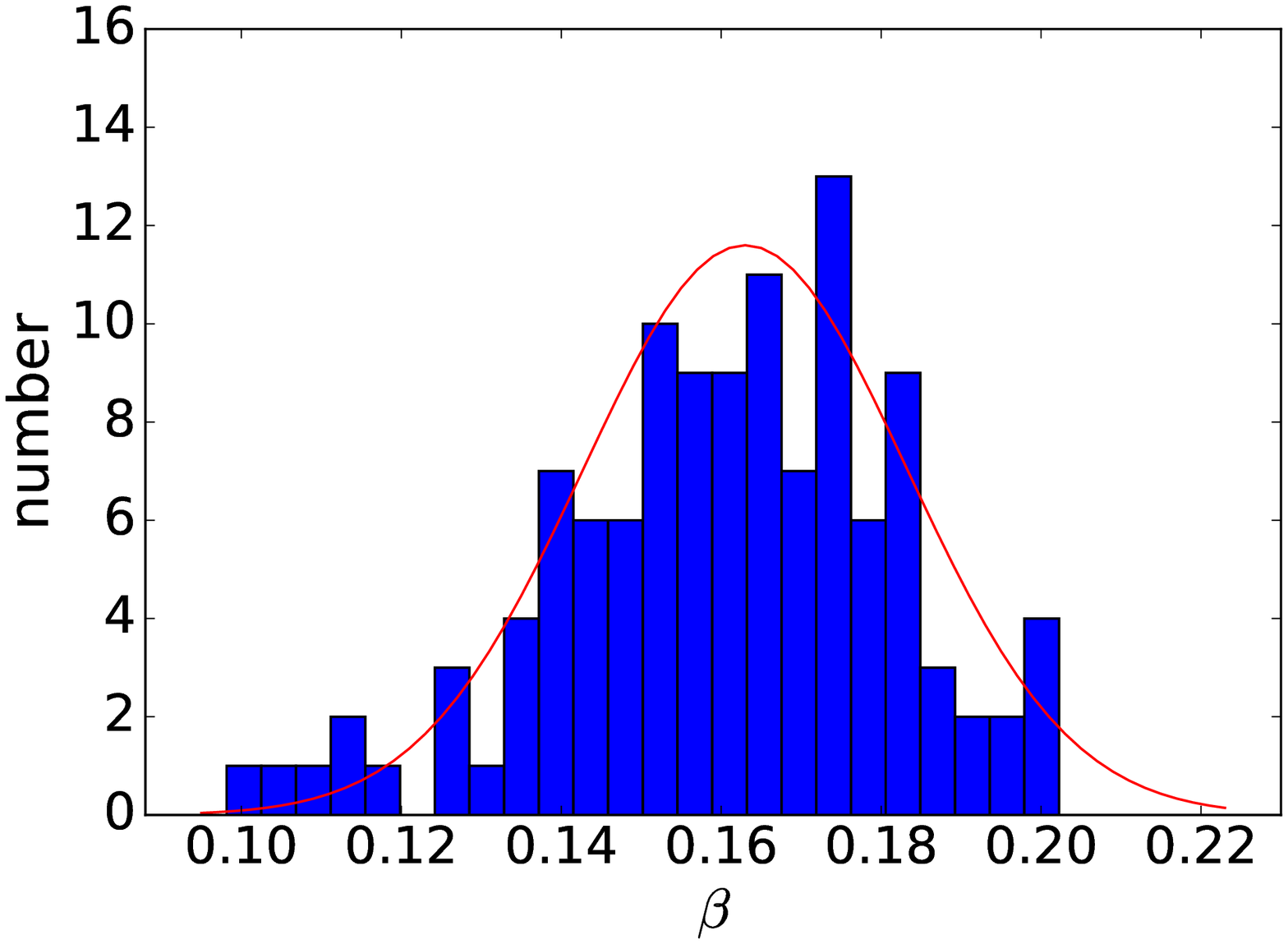}}
\subfigure[]{\label{Fig.kappa_distri}\includegraphics[height=0.3\textwidth, width=0.32\textwidth]{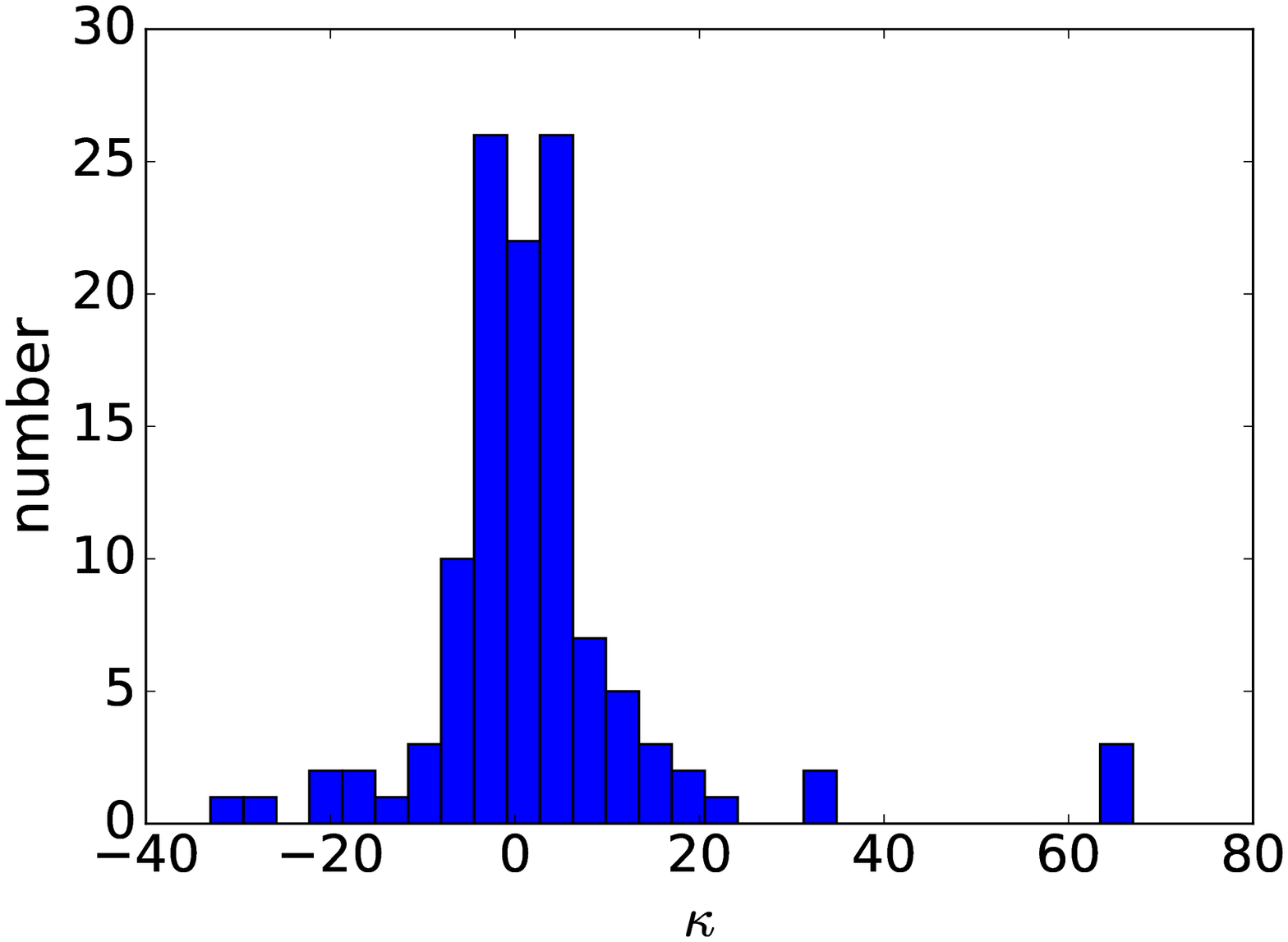}}
\subfigure[]{\label{Fig.lamda_distri} \includegraphics[height=0.3\textwidth, width=0.32\textwidth]{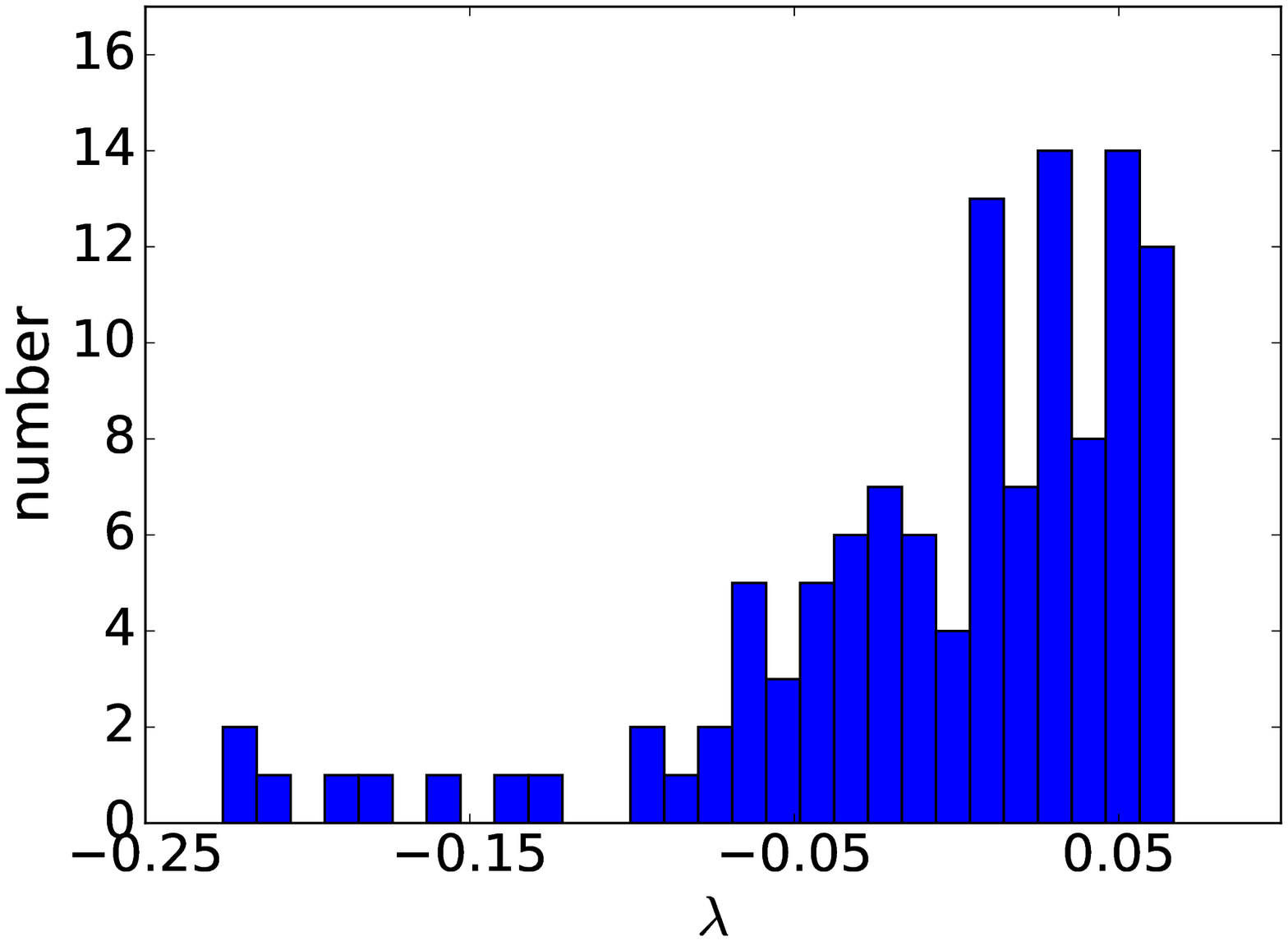}}
\caption{\label{Fig:distri}
The distributions of the Rastall dimensionless parameter $\beta$ (Figure a),  Rastall gravitational coupling constant $\kappa$ (Figure b) and Rastall constant parameter $\lambda$ (Figure c) for 118 galaxies. The red curve in Figure a is the best fitting line with the Gaussian distribution.}
\end{figure*}

In this work, totally 118 gravitational lensing samples are used to constrain $\beta$. The lensing samples come from SLACS \citep{2009ApJ...705.1099A, 2010ApJ...724..511A}, BELLS \citep{2012ApJ...744...41B}, BELLS GALLERY \citep{2016ApJ...833..264S} and SL2S \citep{2013ApJ...777...98S}. For SLACS and SL2S samples, the slope $\gamma$ for each lens galaxy has been provided \citep{2010ApJ...724..511A, 2013ApJ...777...98S}. But for BELLS and BELLS GALLERY samples, we have to use the joint analysis of gravitational lensing and dynamical data \citep{2003ApJ...583..606K, 2004ApJ...611..739T, 2006ApJ...649..599K} to infer the best slope $\gamma$ \citep[the result have been published in][]{2018MNRAS.480..431L}. Next, we use $\gamma$ to obtain the value of $\beta$ for each lens galaxy using Equation \ref{Equ:gamma_beta}, and then plot the distribution of $\beta$ in Figure \ref{Fig.beta_distri}. We find that the distribution of $\beta$ is approximately Gaussian. The mean value of the Gaussian distribution is $\beta=0.163\pm0.001$(68\% CL, the flowing are same), with a minor scatter of $\delta=0.020\pm 0.001$. 

%We find $\beta$ actually acts similar as slope $\gamma$ which determines the mass distribution of ETGs. If assuming $\omega=0$ we can write Equation \ref{equ:density} as
%\begin{equation}
%\rho_{tot}=\rho_{0}(\frac{R}{R_0})^{3-\frac{3\beta}{1-3\beta}}.
%\label{Equ:power_law}
%\end{equation}

After known $\beta$, we can infer the values of $\kappa$ and $\lambda$ from Equation \ref{equ:Newton_limit}. In order to simplify the calculation, we difine $8\pi G\equiv1$, then $\kappa_G=1/2$. Now we obtain
\begin{equation}
\kappa=\frac{4\beta-1}{6\beta-1},    \ \ \ \ \ \ \ \ \ \  \ \ \    \lambda=\frac{\kappa-1}{6\kappa^2-4\kappa}.
\label{Equ:kappa_lambda}
\end{equation}
The distributions of $\kappa$ and $\beta$ are shown in Figure \ref{Fig.kappa_distri} and \ref{Fig.lamda_distri}, respectively. Here, we note that $\beta$ can not be $1/6$, because if $\beta=1/6$, then $\kappa=\pm\infty$. It is not allowed in Equation \ref{equ:rastall_equation} and Equation \ref{equ:Newton_limit}. From Equation \ref{Equ:gamma_beta} we know that if $\beta=1/6$, the mass density slope of ETGs would be $\gamma=2$, which is an isothermal distribution. However, $\beta=1/6$ is not allowed in the Newtonian approximation of Rastall gravity, meaning that the mass distribution of ETGs can not be absolutely isothermal. 

\section{Discussion}
\subsection{The intervals of $\beta$}
The energy condition for the Rastall gravity can help us to exame the rationality of our result for $\beta$. The weak energy condition (WEC) requires that the measured total energy density of all matter fields for any observer traversing a time-like curve is never negative. In the standard locally non-rotating frame \citep[LNRF,][]{ 2015arXiv151201498T}, it requires $T_{\eta\xi}\mu^{\eta}\mu^{\xi}\geq 0$, where $\mu^{\eta}$ is the four velocity and $\mu^{\xi}$ is the time-like vector. In our study, we ignore the pressure of the matter and let the equation of state as $\omega=0$.  Therefore, the WEC requires $\rho\geq 0$. 

%$T_{\mu\nu}\mu^{\mu}\mu^{\nu}\geq 0$, where $\mu^{\mu}$ is the four velocity and $\mu^{\nu}$
%The strong energy condition (SEC) is a geometrical condition arising from the assumption of the convergence of the geodesics at infinity leading to
%\begin{equation}
%R^{\mu\nu}\mu^{\mu}\mu^{\nu}\geq 0,
%\end{equation}
%combined with Equation $\ref{equ:rastall_equation}$, we obtain
%\begin{equation}
%\kappa(T_{\mu\nu}-\dfrac{2\beta -1}{8\beta -2}Tg_{\mu\nu})\mu^{\mu}\mu^{\nu}\geq 0.
%\end{equation}

The attractive nature of gravity implies that the geodesics should converge leading to
\begin{equation}
R_{\eta\xi}\mu^{\eta}\mu^{\xi}\geq 0,
\end{equation}
called the strong energy condition (SEC). It can be combined with Equation \ref{equ:rastall_equation} to reach
\begin{equation}
\kappa(T_{\eta\xi}-\frac{2\beta-1}{8\beta-2}Tg_{\eta\xi})\mu^{\eta}\mu^{\xi}\geq 0,
\end{equation}
which recovers SEC in the framework of GR when $\beta=0$. For a pressureless source, it leads to
\begin{equation}
\frac{\frac{\kappa}{2}(6\beta -1)}{4\beta -1}\rho\geq 0,
\end{equation}
combined with Equation $\ref{Equ:kappa_lambda}$ we obtain the strong energy condition to be $\rho\geq 0$.

Now, combine the requirement of WEC and SEC with Equation \ref{Equ:density}, we finally get
\begin{equation}
\frac{\alpha(1-4\beta)\beta}{\kappa}\geq 0.
\label{Equ:sec_wec}
\end{equation}
In order to infer the admitted intervals for $\beta$ from WEC and SEC, we also need to consider
\begin{equation}
\kappa=\frac{4\beta-1}{6\beta-1}.
\label{Equ:kappa}
\end{equation}
According to Equation \ref{Equ:sec_wec} and \ref{Equ:kappa}, if $\alpha> 0$ and $\kappa>0$, we get $0\leq\beta\leq\frac{1}{6}$. If $\alpha< 0$ and $\kappa<0$, we get $\dfrac{1}{6}\leq\beta\leq\frac{1}{4}$. The general range of the slope $\gamma$ for the inner region of ETGs is $1<\gamma<3$, corresponding to $0<\beta<2/9$, satisfying the above two cases. Other two cases, $\alpha > 0$ and $\kappa <0$, or $\alpha < 0$ and $\kappa >0$, would result in $\beta\leq 0$ or $\beta\geq\frac{1}{4}$, which is not allowed in our study. In addition, if $\alpha=0$, $\kappa$ can be both positive and negative. 

Rastall gravity has also been applied to study some other intresting problems, and these works also gave some different intervals for $\beta$. For example, \cite{2017CaJPh..95.1257M} studied the properties of traversable wormholes in the framework of Rastall gravity. They indicated that for some values of $\beta$, unlike the Einstein theory, traversable wormholes were also supported by the ordinary matter in the Rastall theory. They found that if the equation of state $\omega\rightarrow 0$, to allow the existence of wormholes, the Rastall dimensionless parameter should be $\beta>1/2$. Our study for ETGs is also the case of $\omega\rightarrow 0$, the general range of $\beta$ is $0<\beta<2/9$. Therefore, the special type of traversable wormholes ($\omega\rightarrow 0$ case) studied in \cite{2017CaJPh..95.1257M} can not be supported by the allowed interval of $\beta$ obtained in our study. Another interesting inference from Rastall gravity is that the coupling between geometry and pressureless matter fields can theoretically describe the current accelerated and inflation phases of the universe without introducing strange sources (e.g., dark energy) \citep{1996PhLB..366...69A, 1996GReGr..28..935A, 2003JCAP...05..008A, 2017PhRvD..96l3504M,  2017EPJC...77..259M}. In particular, in \cite{ 2017PhRvD..96l3504M}, the expansion of the universe requires $\beta<1/4$ or $\beta>1/3$. Although our result satisfies the case of $\beta<1/4$,  we do not think it could be a supportion for that the coupling between geometry and pressureless matter fields in Rastall grivity can explain the expansion of the universe. Because our results are just limited to the galaxy scale. For a different scale (e.g., galaxy cluster scale or cosmical scale), the observational values for $\beta$ may be different.

%Our result of $\beta$ (roughly $0<\beta<2/9$) satisty the requirement of the accelerating phase of the universe.

\subsection{The power-law mass density profile}
The well known assumption of power-law mass distribution for ETGs has not been explained with GR. Now, we find that it is a nature result of Rastall gravity.  The Rastall dimensionless parameter $\beta$ in Rastall grivity determines the mass distribution of ETGs. At beginning, the power-law mass-density profile was used in \cite{2003ApJ...583..606K}. In their work, they supposed the total mass distribution of luminous plus dark matter of ETGs follows a single power-law $\rho_{tot}\propto r^{-\gamma}$, where $\gamma$ is the ``effective slope". Then they used this profile to fit the lensing system 0047-281 and got $\gamma=1.90\pm0.05$. \cite{2004ApJ...611..739T} and \cite{2006ApJ...649..599K} made further discussions for the rationality of the power-law mass distribution assumption. They suggested that if the mass-density profile of ETGs are different from a power-law, one should expect the density slope inside Einstein radius to change with the ratio of Einstein radius to effective radius. However, their works, as well as the later works \citep[e.g.,][]{2009ApJ...703L..51K, 2011ApJ...727...96R, 2013ApJ...777...98S}, found that the mass-density slope has no relation with the ratio of Einstein radius to effective radius, supporting that the assumption of a single power-law shape for the total mass-density profile is valid. All of these works indicated that the assumption of the power-law mass-density profile for the inner region of ETGs seems to be reasonable, but no one has theoretically explained it. Now we present that the power-law total mass-density profile is a natural result of Rastall gravity. 

In fact, our work is not the first work to study gravitational lensing effects in the Rastall gravity framework.  \cite{2001Ap&SS.278..383A} and  \cite{2005Ap&SS.298..519A} have studied the  lensing probability in a flat cosmological model with Rastall gravity. They found that the lensing probability predicted by Rastall gravity is consistent with the gravitational lensing data. The Rastall gravity could also be used to study other gravitational lensing problems, such as the strong or weak lensing  caused by galaxy cluster, the weak lensing caused by large-scale structure. Next, we will use the Rastall gravity to study the galaxy cluster scale weak lensing. We also expect that this kind of weak lensing could provide some constrains to the Rastall parameters.

%However, under the perfect fluid assumption, we also find an isothermal mass distribution for ETGs is not allowed in Rastall gravity. 

\section{Conclusions}
In this paper, we derived a power-law mass-density profile for ETGs from Rastall gravity under the assumptions of spherical symmetry mass distribution and the perfect fluid matter. Then we use 118 galaxy-galaxy strong gravitational lensing systems to constrain the parameter $\beta(=\kappa\lambda)$ in Rastall gravity and obtain $\beta=0.163\pm0.001$(68\% CL) with a minor intrinsic scatter of $\delta=0.020\pm 0.001$.  Our results satisfy both the strong energy condition and the weak energy condition.  The Rastall dimensionless parameter $\beta$ determines the mass distribution of ETGs, acting similarly as the experiential power-law mass density slope $\gamma$ \citep{2003ApJ...583..606K, 2006ApJ...649..599K}.  We also find that an absolutely isothermal mass distribution for ETGs is not allowed in the the framework of Rastall gravity.

\section{Acknowledgements}
We thank the anonymous referee for the constructive comments and suggestions that significantly improved this paper. We acknowledge the financial support from the National Natural Science Foundation of China 11573060 and 11661161010.

\end{document}